\documentclass[twocolumn,nofootinbib]{revtex4}
\usepackage{amsmath,graphicx}
\begin{document}

\newcommand{\myunit}[1]{\textup{#1}}
\newcommand{\zu}[1]{\textup{#1}}
\newcommand{\kgunit}{\myunit{kg}}
\newcommand{\gunit}{\myunit{g}}
\newcommand{\munit}{\myunit{m}}
\newcommand{\sunit}{\myunit{s}}
\newcommand{\degunit}{\ensuremath{\,^{\circ}}}
\newcommand{\degcunit}{\degunit\textup{C}}
\newcommand{\junit}{\myunit{J}}
\newcommand{\nunit}{\myunit{N}}
\newcommand{\kunit}{\myunit{K}}
\newcommand{\unitdot}{\!\cdot\!}
\newcommand{\vc}[1]{\textbf{#1}}

\title{Comment on an Effect Proposed by A.G.~Lebed in Semiclassical Gravity}
\author{B.~Crowell}
\affiliation{Natural Science Division, Fullerton College, 321 E.~Chapman Ave., Fullerton, CA 92832}

\date{\today}

\begin{abstract}
  A.G.~Lebed has given an argument that when a hydrogen atom is transported slowly to a different
  gravitational potential, it has a certain probability of emitting a photon. 
  He proposes a space-based experiment to detect this effect.
  I show here that his arguments also imply the existence of nuclear excitations, as well
  as an effect due to the earth's motion in the sun's potential.
  This is not consistent with previous results from underground radiation
  detectors. It is also in conflict with astronomical observations.
\end{abstract}

\maketitle

\section{Introduction}

\newcommand{\cl}{ref.~\cite{lebed}}

A.G.~Lebed\cite{lebed} has calculated the behavior of a hydrogen atom that is prepared in its
ground state and then transported slowly through the earth's gravitational field.
The metric is taken to be
\begin{equation}\label{eqn:lebed-metric}
  ds^2=(1+2\phi)dt^2-(1-2\phi) d\ell^2,
\end{equation}
where $c=1$, the spatial line element is $d\ell^2=dx^2+dy^2+dz^2$, and the gravitational potential $\phi\ll 1$ varies
between the initial and final positions.
In \cl, the metrical dilation
is treated as an adiabatic perturbation of the Hamiltonian. This perturbation puts the atom into
a superposition of states, leading to decay by photon emission with probability $P$.

The more standard approach in semiclassical
gravity\cite{wald} would have been to rewrite the Einstein field equations $G_{ij}=8\pi T_{ij}$ as
$G_{ij}=\langle 8\pi T_{ij}\rangle$. In that approach, any excitation of the atom by gravity
would result from the curvature of spacetime, which depends on the second derivatives of the  metric.
In order to calculate the curvature we would need to know not just a
parametrization of the form \eqref{eqn:lebed-metric} but more detailed information about how the metric varied
in a neighborhood of the atom's world-line. 
But the phenomenon predicted in \cl\  is not a curvature
effect. It is an effect of the metric itself (not its derivatives) on
an arbitrarily small particle. Furthermore, in the adiabatic approximation
the excitation probability $P$ depends only on the
value of the metric tensor at the final position as compared with its initial value.
Such a comparison is coordinate-dependent, and is presumably meant to be carried out
in harmonic coordinates. To a local observer,
the emission of radiation would appear to violate conservation of energy; this would have
to be accounted for somehow by the extraction of energy from the gravitational field.

Generalizing to systems bound by forces other than electromagnetic ones, 
let the potential be $U=-kr^n$. This allows us to recover the case of an electrical attraction when
$k=e^2$ and $n=-1$, but we can also mock up a bag of quarks by taking $n=1$.
Carrying through the calculations of \cl\  with these generalizations, we obtain a perturbation to the
Hamiltonian
\begin{equation}
  \Delta H = (m+K+U)\phi+V\phi,
\end{equation}
where the factor in parentheses is subsumed in the particle's mass,
and the second term contains the operator $V=2K-nU$, which has a vanishing expectation value by 
the quantum virial theorem.\cite{carlip}
The probability of excitation is the product of two factors, which I notate as
\begin{equation}\label{eqn:prob}
  P=f_g^2 f_s^2.
\end{equation}
The first of these is gravitational, 
\begin{equation}
  f_g=\phi-\phi_0.
\end{equation}
The second one depends on the structure of the quantum-mechanical system, and is given by
\begin{equation}\label{eqn:fs}
  f_s = \langle 2|V|1\rangle/\Delta E,
\end{equation}
where $\Delta E=E_2-E_1$ is the difference in energy between the two states.
The factor $f_s$ is of order unity for hydrogen.\cite{lebed}

\section{Solar Effect}

We should have effects not just from the earth's gravity but from the sun's as well. 
Moving an atom between the earth's surface and a distant
point, as originally proposed, produces $f_g=6.9\times10^{-10}$. The earth's motion from
perihelion to aphelion gives $f_g=3.3\times10^{-10}$. Since these are within a factor of 2 of one
another, it would seem that given any space-based experimental design, one could achieve a far greater
sensitivity at a much lower cost by substituting a large sample of atoms on earth for a small
sample aboard a space probe.


\section{Scale-Independence, and Some Systems of Interest}

The only dependence of the predicted effect on the structure of the system comes from the value of
the exponent $n$ in the potential, and from
equation \eqref{eqn:fs} for the quantity $f_s$. The latter is the dimensionless ratio
of two energies, and does not depend on the linear dimensions of the system,
so that the effect is completely independent of scale. This scale-independence
arises because the effect is due to the gravitational potential at a point,
rather than to curvature.
We therefore expect generically
that excitations would occur for almost any quantum-mechanical system, including
hadrons, nuclei, atoms, and molecules of all sizes.

The structure factor $f_s$ depends on
the matrix element  $V_{21}=\langle 2|V|1\rangle$, and calculating this
matrix element explicitly for all of these systems would be a significant project.
But regardless of the value of the exponent $n$ in the potential,
$V_{21}$ is essentially a measure of the total internal energy of the system,
so it is not unreasonable, for the sake of some order-of-magnitude estimates, to
assume that it can be estimated as such. There is a selection
rule that there can be no change in spin or parity. We would also like to find
states 1 and 2 such that 2 has the dynamical character of a radial excitation of 1 
so that there is a significant coupling to the metrical dilation
measured by $V$, and so that a
many-body system can be treated by taking $r$ to be a generalized coordinate
describing a collective spatial dilation. Under these assumptions
we have simply $f_s\sim E/\Delta E$. This ratio will often be of order unity,
as in hydrogen, but in some cases it is orders of magnitude higher.

\begin{table}[h]
\centering
\begin{tabular}{|l|p{20mm}|p{20mm}|l|p{15mm}|}
\hline
 & system & excitation & $E/\Delta E$ & radiation \\
\hline
i & proton & $\text{N}(1440)\text{P}_{11}$ & $\sim1$ & $\pi^0$ \\
\hline
ii & heavy nucleus & isoscalar giant monopole resonance & $\sim10^2\text{--}10^4$ & particle emission \\
\hline
iii & H atom & $n=2$ & $\sim1$ & photon \\
\hline
iv & $\text{C}_{60}$ (fullerene) & $0.061\ \text{eV}$ ``breathing'' mode & $\sim10^6$ & photon \\
\hline
\end{tabular}
\caption{\label{table:systems}Properties of some relevant quantum-mechanical systems.}
\end{table}

Table \ref{table:systems} summarizes some systems in which one would expect the effect
to occur. These are labeled i through iv for reference. The excitations in systems
i, ii, and iv have all been interpreted as ``breathing'' modes of vibration, which
suggests that the estimate $f_s\sim E/\Delta E$ is reasonable for them.

For an atomic or molecular experiment of the type suggested in \cl, it is to be remarked
that rather than hydrogen, other systems, such as large molecules, iv, would be expected to produce excitations
with probabilities greater by a factor of $f_s^2=10^{12}$ or more. Molecules of this size
have been diffracted by a grating in experiments, which demonstrates the possibility of
placing them in a superposition of states.\cite{arndt}

The transition rate for example ii is not straightforward to estimate by the techniques used here, but
it raises the possibility that
otherwise stable nuclei on earth would decay by particle emission.

\section{Memory Effect}\label{sec:memory}

In the simple example of constant-velocity motion through a uniform gravitational field,
the effect is predicted to grow quadratically with the time since the system was first formed.
Such a non-exponential decay means that a hydrogen atom, for example, would retain
a memory of the potential in which it was formed, and that an observer could
access this memory, at least at the statistical level.

Although \cl\ proposes using a ``tank of a pressurized hydrogen'' aboard a satellite,
it seems likely that a collision would erase a hydrogen molecule's memory. It would therefore
be preferable to work with a system such as a nucleus, which may remain isolated from its environment
for billions of years.

\section{Hadronic Excitations}\label{sec:nuclear}

In the remainder of this paper I will consider excitation of the proton, example i
in table \ref{table:systems}. There is an excited state, labeled $\text{N}(1440)\text{P}_{11}$,
that matches the spin-parity $1/2^+$ of the ground state and is believed to be
a radial excitation of the three quarks. For these reasons, we expect $f_s\sim E/\Delta E\sim 1$
for excitation of this state. Its most frequent mode of decay is radiation of a neutral pion.

Let us estimate the rate at which hydrogen nuclei on earth would be expected to emit pions.
The rate of decay is
\begin{align}
  \Gamma &= dP/dt \\
         &= f_s^2 d(f_g^2)/dt \\
         &= 2 f_s^2 (\phi-\phi_0)d\phi/dt \label{eqn:decay-rate} \\
         &= -2 f_s^2 (\phi-\phi_0)\vc{g}\cdot\vc{v},
\end{align}
where $\vc{g}$ is the sun's gravitational field and $\vc{v}$ is the velocity of a hydrogen
atom, both as measured by a static observer. (The dependence of the effect on $\vc{g}$ violates
the equivalence principle, and \cl\ explicitly interprets the effect as such a violation.)

The potential $\phi_0$ would be the potential at which the proton was formed, and we need to
define this time of formation. As discussed in section \ref{sec:memory},
a memory of this potential is carried by the system in this theory, and it is not entirely
obvious what would serve to erase its memory. I will assume that this occurs when there is
any collision, i.e., when the proton approaches another nucleus, coming within the range of the
strong nuclear force. Some protons on earth were formed during big-bang nucleosynthesis (BBN), but
others have participated in collisions at the cores of stars.
Still others will have undergone their last collision during a type I supernova, near
the surface of a white dwarf star.
It is a consequence of the model in \cl\ that all these classes of protons differ
in their subsequent behavior. Since \cl\ assumes a static spacetime, we will not consider the
BBN component, which has existed over cosmological timescales. For the component whose
memory was reset by collisions at the core of a star,
we estimate $|\phi_{0,c}|\sim 3M_\odot/R_\odot\sim6\times10^{-6}$,
which is $\ll 1$ as assumed in \cl, but much larger than the changes in potential considered there.
For those that have been recycled through a supernova, we may have $|\phi_{0,sn}|\sim 10^{-4}$,
but the astrophysics is more complicated in this case, so in the following discussion
I will use the more conservative estimate based on $\phi_{0,c}$. 
(Variations in the gravitational potential within the galaxy are smaller than these values,
with the potential experienced by our solar system being about $-1.7\times10^{-6}$.\cite{kafle})
In section \ref{subsec:g2},
I give an astronomical check on the predictions of \cl\ that is entirely independent of such
estimates of $\phi_0$, or of the assumption that the memory of $\phi_0$ can be retained for
long periods of time.

Due to the sign of $\phi-\phi_0$, these protons would radiate during the half of the year when
the earth is moving from perihelion to aphelion. The probability of decay during such a six-month
period is estimated to be $7\times10^{-15}$, resulting in an average rate of radiation
\begin{equation}\label{eqn:lebed-pred}
  \Gamma\sim 4\times10^{-22}\ \sunit^{-1}.
\end{equation}
In the following section I will compare this with experiment.

\section{Empirical Bounds}
\subsection{Underground Radiation Detectors}
The excited state of the proton described above decays by emission of a pion, which
would be detectable by its decay into a high-energy electromagnetic cascade.
A number of underground experiments have already been carried out to detect neutrinos or search
for dark matter, and these experiments would have been extremely sensitive to such
a phenomenon. One of the most important sources of background in such an experiment is
cosmic-ray-induced muons, which also create high-energy cascades. This is the reason
that the experiments are carried out underground. In addition, various steps are taken
in order to block or reject the resulting events, including the rejection of the kind
of high-multiplicity, high-energy events that are of interest here. However, preliminary
studies have been carried out in which the experimental trigger was left wide open, in order
to precisely characterize the muon-induced background.

I consider here the Large-Volume Detector apparatus (LVD), which consists of 983 tons of hydrocarbon
scintillator located about one kilometer underneath the Gran Sasso massif. This scintillator contains 
$N=9.4\times10^{31}$
hydrogen atoms. Multiplying by the rate of radiation estimated in section \ref{sec:nuclear}, we find a predicted count rate
\begin{equation}
  R =\Gamma N \sim 2\times10^{10}\ \sunit^{-1}.
\end{equation}

The LVD collaboration has made detailed measurements of the flux of muons through
their detector.\cite{aglietta}\cite{aglietta2}
In these observations, the direction from which the muon entered the detector is reconstructed,
and is found to decay exponentially with the depth of rock through which the muon would have had to travel.
For angles very close to horizontal, the flux of muons is measured to be 
$\sim 10^{-12}\ \zu{cm}^{-2}\sunit^{-1}\zu{sr}^{-1}$.

The effect discussed here would have caused the creation of similar high-energy electromagnetic cascades
originating from protons within the detector's own active volume. These cascades would have been
emitted isotropically, and therefore would have shown up as part of the same count rate attributed above
to muon-induced cascades. To be consistent with the measurement above, their rate would have been limited to
\begin{equation}
  R \lesssim 3\times 10^{-5}\ \sunit^{-1}.
\end{equation}
This is inconsistent with expectations from \cl\ by a factor of $10^{15}$.
Even this is likely to be an underestimate; references \cite{aglietta} and \cite{aglietta2}
do not state explicitly whether the analysis looked for such events coming in the upward direction, but if
so, then the effect implied by \cl\ would be ruled out by even more orders of magnitude.

The Super-Kamiokande collaboration has searched for proton decays\cite{kamiokande} via the processes
$\zu{p}\rightarrow\zu{e}^+\pi^0$ and $\zu{p}\rightarrow\mu^+\pi^0$, finding a limit on the rate of
decay of $\sim 10^{-41}\ \sunit^{-1}$. Since the analysis of the data from this experiment employed
sophisticated kinematic reconstructions in order to search for these specific processes, this rate
cannot be directly compared with the estimate in equation \eqref{eqn:lebed-pred} for
the rate of $\zu{p}\rightarrow\zu{p}\pi^0$, which would have shown up in the analysis as
an event that violated conservation of energy-momentum. Nevertheless,
since the two rates differ by a factor of $10^{20}$, it appears unlikely that a background
due to neutral pion emission would have gone unnoticed.

\subsection{The Galactic Center}\label{subsec:g2}
An independent empirical check on the predicted effect comes from observations of stars and gas clouds
in tight orbits about Sagittarius A*, the supermassive black hole at the center of
our galaxy.
Because the potentials experienced by these objects are much larger than
the value of $\phi_{0,c}$ assumed above, the behavior of the protons in this environment is independent of
the assumption employed earlier that a proton retains a memory of the gravitational potential over billions of years.

\begin{figure}
\includegraphics{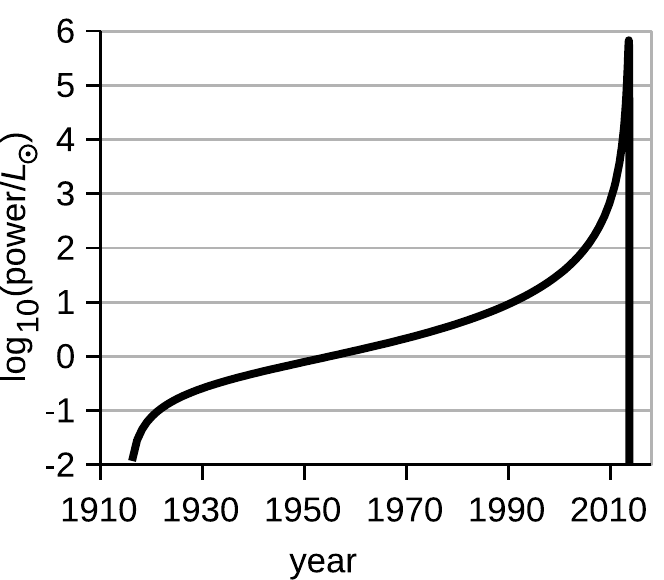}
\caption{\label{g2} Simulation of the instantaneous power liberated by pion emission in the gas cloud G2, from
apastron to periastron. $M=2M_\odot$, $\phi_0=0$, and $f_s=1$ are assumed.}
\end{figure}

The object G2 is in a highly elliptical, high-velocity
orbit about Sag A*.
It has been interpreted as a cloud of gas and dust which is heated by a star
hidden at its center.\cite{witzel} Its orbit,\cite{gillessen2} with an eccentricity of 0.966,
is nearly a radial free-fall toward A*, with periastron taking place in 2013. Its motion and magnitude
in the Ks band ($2.2\ \mu\munit$) were observed from 2005 to 2014, and its brightness remained
constant throughout this period ($\lesssim 0.2$ magnitudes of variation).
A fit to a blackbody curve gives a luminosity of $29L_\odot$.
If the central star is on the
main sequence, then its mass is $2M_\odot$; if not, then the mass may be smaller.
Figure \ref{g2} shows the rate at which energy would have been liberated through pion emission
by a body of mass $2M_\odot$, composed of hydrogen, tracing out G2's Keplerian orbit.
The result is a spike in power equivalent to $10^6L_\odot$, which is not consistent with
the observed lack of variation, and indeed probably would have destroyed the star.
The peak power is not appreciably changed by
changing the potential of formation from zero to
$\phi_0=-8\times10^{-6}$, which is a typical potential experienced by G2 during its orbit.
Therefore this conclusion remains valid even if a proton's memory of its potential of formation
is limited to decades rather than billions of years.

Another astronomical test is available from observations of the star S2, a main-sequence
B1 star that is also orbiting Sag A*. Assuming $M=15M_\odot$, $\phi_0=0$, and $f_s=1$,
a calculation similar to the one described above gives a peak power of $2\times10^7L_\odot$,
or about 700 times the normal luminosity of a B1V star.
Intriguingly, observations have shown an unexpected 
40\%  rise in the star's luminosity when it was
near perihelion, but explanations have been suggested using standard physics.\cite{gillessen}

\section{Conclusions}

I have examined a provocative claim by A.G.~Lebed that
a quantum-mechanical system can be induced to emit radiation by moving it
slowly to a different gravitational potential. This prediction is incompatible
with existing terrestrial experiments and astronomical observations.

\section{Acknowledgments}
I thank B.~Shotwell and S.~Carlip for helpful discussions.

\bibliography{lebed}

\end{document}